\documentclass[preprint,showpacs,preprintnumbers,amsmath,amssymb]{revtex4}

\usepackage{amsmath,amssymb}
\usepackage{graphicx} 
\usepackage{dcolumn} 
\usepackage{bm} 

\begin{document}

\title{\bf Spin Bias Measurement Based on a Quantum Point Contact}

\author{Yanxia Xing}
\author{Qing-feng Sun}
\email{sunqf@aphy.iphy.ac.cn} \affiliation{Institute of Physics,
Chinese Academy of Sciences, Beijing 100080, China.}%
\author{Jian Wang}
\affiliation{ Department of Physics and the center of theoretical
and computational physics, The University of Hong Kong, Hong Kong,
China.}

\begin{abstract}
Electron charge transport through a quantum point contact (QPC)
driven by an asymmetric spin bias (SB) is studied. A large charge
current is induced when the transmission coefficient of the QPC
jumps from one integer plateau to the next. Furthermore, for an open
external circuit, the induced charge bias instead of the charge
current is found to be quite large. It provides an efficient and
practical way to detect SB by using a very simple device, a QPC or a
STM tip. In addition, with the aid of magnetic field, polarization
direction of the SB can also be determined.
\end{abstract}

\pacs{85.75.-d, 85.30.Hi, 73.63.Rt, 73.23.-b}
\maketitle

The field of spintronics has received more and more attention
recently. The spin current is one of the most important physical
quantities in spintronics,\cite{spintronics,ref2} similar to the
role of the charge current played in electronics. Important issues
of spintronics include how to generate, manipulate, and detect the
spin current. Various methods have been proposed to generate spin
current.\cite{theory,experiment,optical,mag,isH,sH} Up to now, the
spin current has been generated by various means, e.g., the pump
excitation,\cite{experiment} the optical injection,\cite{optical}
the magnetic injection\cite{mag,isH}, and the spin Hall
effect.\cite{sH} To measure the spin current, the light-emitting
diode or Kerr-rotation spectroscopy has been experimentally used to
detect the spin-current induced spin accumulation near the
boundaries of the sample.\cite{sH} Moreover, an electric measurement
was also realized via the inverse spin Hall effect.\cite{isH} In
addition, theoretical proposals on indirect measurement of spin
current were also available.\cite{ref9,ref11,ref12,aref1} These
include the detection of the spin torque caused by a spin current
flowing through a ferromagnetic-nonmagnetic interface\cite{ref9} and
the detection of the spin-current induced electric
field.\cite{ref11,ref12} However, all these experimental instruments
are delicate and complicated. They always involve the optical and
magnetic factors, or spin-orbit interaction. Up to now, there is yet
a practical and effective approach to measure spin current. Hence
the measurement of the spin current remains a challenge.

In charge transport, one can measure the charge bias instead of the
charge current. For a spin current flowing through a device, a
spin-dependent chemical potentials (spin bias, SB) is usually
induced that is the driving force of the spin current. We can
measure this SB instead of spin current. In this paper, we propose
an effective method to detect the SB by using a quantum point
contact (QPC) or a STM tip. The QPC is the simplest device in
mesoscopic physics. The transport property of the QPC has been
investigated extensively. Its conductance versus the gate voltage
$V_g$ shows a series of step structures at the value $2 n e^2/\hbar$
($n=1,2,...$). Experimentally, the QPC has been used as a charge
sensing detector to reliably probe the number of electrons in the
quantum dot.\cite{ref13} Our results show that a charge bias or
charge current emerges when the QPC is biased by an asymmetric
SB.\cite{ref15} Therefore by measuring the induced charge bias, we
can detect the SB in a very simple way.

The proposed device is a QPC device under an asymmetric SB as shown
in Fig.1a.\cite{note1} Our task is to ``experimentally'' measure the
SB. A concrete example of the device is shown in Fig.1c that is
based on the experimental setup used in Ref.\cite{isH} where a
ferromagnetic (FM) lead crossed over an aluminium (Al) lead. A
charge current is injected from the FM lead into one terminal of Al
lead, then a pure spin current is generated and flows into another
terminal of the Al lead. A spin-dependent chemical potentials (i.e.
SB) is created there.\cite{isH} Of course, the SB can also be
generated using other methods. In this paper, we propose to measure
this SB by using a STM tip (see Fig.1c) or a QPC device.

Let us first discuss the working principle of detecting the SB using
the QPC.
Due to the asymmetric SB $V_s$,\cite{ref15} the chemical
potentials on the left lead are spin-dependent with
$\mu_{L\uparrow}= E_f + V_s$ and $\mu_{L\downarrow}= E_f - V_s$, but
the chemical potentials on the right lead are still spin independent
with $\mu_{R\uparrow}= \mu_{R\downarrow}= E_f$ (see Fig.1a). Under
the SB $V_s$, the charge current $I_{\sigma}$ (with
$\sigma=\uparrow,\downarrow$) is given by the
Landauer-B$\ddot{u}$ttiker formula:
\begin{equation}
I_{\sigma}=\frac{e}{h}  \int dE~ T(E)[f_{L\sigma}(E)-f_{R\sigma}(E)]
\label{Jc}
\end{equation}
where $T(E)$ is the transmission coefficient and $f_{p\sigma} =
1/\{\exp[(E-\mu_{p\sigma})/k_B {\cal T }] +1 \} $ is the Fermi
distribution of the leads. From Eq.(1), the spin-up and spin-down
currents, $I_{\sigma}$ with $\sigma=\uparrow , \downarrow$ or
$\sigma=\pm$ are mainly determined by $T(E)$ with $E$ between
$\mu_{R\sigma}=E_f$ and $\mu_{L\sigma}= E_f + \sigma V_s$. Note that
the transmission coefficient $T(E)$ is energy dependent. In
particular, when the energy $E$ is at the middle of the jump as
shown in Fig.1b, $T(E)$ is strongly dependent on $E$. In general,
$|I_{\uparrow}|$ is not equal to $|I_{\downarrow}|$. As a result, a
net charge current $I_c = I_{\uparrow}+I_{\downarrow}$ occurs
although $I_{\uparrow}$ and $I_{\downarrow}$ flow in opposite
directions. So by measuring the induced charge current $I_c$, we can
detect the SB $V_s$. In the following, we shall investigate $I_c$ as
well as the relation between $I_c$ and $V_s$ in detail.

To calculate $I_c$, we assume the QPC device is described by the
Hamiltonian $H = (p_x^2 +p_y^2)/2m^* +V(x,y)$. In the QPC region
($|x| \leq L/2$ and $|y| \leq W/2$), the potential is assumed in a
saddle form $V(x,y)=\beta\frac{y^2}{W^2} -\frac{V_g}{{\rm
cosh}^2(\alpha x/L)^2}$. In the lead regions ($|x|>L/2$ and $|y|
\leq W_{lead}/2$), $V(x,y)=0$. $V(x,y)=\infty$ outside of lead and
QPC regions. In the tight-binding representation, the Hamiltonian
becomes:\cite{Hami}
 $H=-t \sum\limits_{<{\bf i}, {\bf j}>,\sigma}
 a_{{\bf i} \sigma}^\dagger  a_{{\bf j}\sigma}
 + \sum\limits_{{\bf i},\sigma}(4t+V_{\bf i})
 a_{{\bf i} \sigma}^\dagger  a_{{\bf i}\sigma}$,
where ${\bf i}= (i_x, i_y)$ labels the site index, $t=\hbar^2/2m^*
a^2$ is the nearest neighbor hopping matrix element, $a$ is the
distance between two neighboring sites, and $V_{\bf i}= V(i_x a, i_y
a)$. From the discretized Hamiltonian $H$, the transmission
coefficient $T(E)$ is given by $T(E)=Tr[{\bf \Gamma}_{L}{\bf
G}^r{\bf \Gamma}_{R}{\bf G}^a]$ where we have used the Green
function ${\bf G}^r(E) =[{\bf G}^a]^{\dagger}=\{E{\bf I}-{\bf
H}_0-\sum_{p}{\bf \Sigma}^r_{p}\}^{-1}$ and the line-width function
${\bf \Gamma}_{p}=i({\bf \Sigma}_{p}^r-{\bf
\Sigma}_{p}^{r\dagger})$. Here ${\bf H}_0$ is the Hamiltonian of the
QPC's region and ${\bf \Sigma}_{p}^r$ is the retarded self-energy
due to the coupling between the lead $p$ and the QPC. With $T(E)$,
the charge current can be calculated from Eq.(1).

In the numerical calculation, we set the hopping strength $t=1$ as
the energy unit. The width of the leads is set to $W_{lead}=500a$,
and the QPC sizes are chosen as $W=50a$ and $L=100a$. The parameters
$\alpha, \beta$ of the QPC's saddle potential are $\alpha =5$ and
$\beta =4t$. The Fermi energy $E_f$ is fixed at $0.6t$ that is near
the band bottom $0$. Fig.2(a) and (b) show the charge current $I_c$
versus the gate voltage $V_g$ at different SB $V_s$ and at different
temperatures ${\cal T}$, respectively. For comparison, the
transmission coefficient $T(E)$ at $E=E_f$ versus $V_g$ is also
plotted in Fig.2(b) (see gray solid line) which exhibits a series of
step structures. For parameters we used, the curve of $T(E_f)$-$V_g$
is very similar to the curve of $T(E)$-$E$ at fixed $V_g$ (not shown
here). In the system, the charge current $I_c$ indeed is nonzero for
nonzero SB $V_s$. From Fig.2 we see that $I_c$-$V_g$ curves exhibit
a series of peaks with approximately the same peak height. In
addition these peaks are well correlated with $dT/dV_g$. For
instance, $I_c$ reaches the maximum value whenever the transmission
coefficient $T$ shows a jump. With increasing of $V_s$, the current
$I_c$ increases as a whole (see Fig.2a). When increasing the
temperature ${\cal T}$ from zero, $I_c$ near the peak decreases
while $I_c$ near the valley increases. As a result, the curve of
$I_c$ versus $V_g$ is smoothed out by temperature effect.

Experimentally, it is more convenient to measure the voltage. In the
following, we consider an open external circuit where a charge bias
$V_{c}$ is induced instead of the charge current $I_c$ passing
through. Due to the induced charge bias $V_c$, the chemical
potential $\mu_{R\sigma}$ of the right lead is shifted from $E_f$ to
$E_f+V_c$. With condition $I_c=0$ in the open circuit, $V_{c}$ can
be determined through Eq.(1). In Fig.3a and b, the ratio $V_c/V_s$
are plotted with the same parameters used in Fig2a and b. We see
$V_c/V_s$ has similar characteristics as $I_c$: $V_c/V_s$ exhibits a
series of peaks, increases with increasing $V_s$, and it is also
smoothed out by the temperature effect. However, there are two big
differences between $I_c$ and $V_c$. First, the peak heights of
$V_c/V_s$ decrease as $|V_g|$ decreases, while they are nearly same
for $I_c$. Second, on the left side of the first peak (at $V_g
\approx -0.56$), $I_c$ is very small but $V_c/V_s$ is always quite
large.

Fig.3c shows the ratio $V_c/V_s$ versus $V_s$ at different gate
voltages $V_g$. When $V_s$ increases from the zero, $V_c/V_s$ rises
quickly and then saturates eventually. We emphasize that the ratio
$V_c/V_s$ is very sensitive to $V_s$ at small $V_s$. For instance
$V_c/V_s$ is larger than 0.1 at $V_s=0.002t$. Fig.3d gives the
effect of temperature on the bias ratio $V_{c}/V_s$. We see that the
peak of $V_c/V_s$ decreases while the valley of $V_c/V_s$ increases
with the temperature. In the high temperature limit, $V_c/V_s$
converge to a constant value for all $V_s$. Notice that the high
temperature limit of $V_c/V_s$ is quite large ($\approx 0.1$). This
means that the SB $V_s$ can be measured even at high temperatures.

Let us discuss the feasibility of our proposal. First, the induced charge
bias $V_c$ is quite large. $V_c/V_s$ can be
over $0.5$ at the certain region (e.g. while $V_g<-0.57$ as shown in
Fig.3). In particular, $V_c/V_s$ is larger than $0.01$ for almost
all regions of the parameter including the SB $V_s$, the gate
voltage $V_g$, the temperature ${\cal T}$, etc. In the present
technology, the charge bias of order of $0.1nV$ can easily be
measured.\cite{ref17} Therefore, our QPC device can detect $V_s$ if
it is over $10nV$. In fact, in the Ref.\cite{isH} the SB $V_s$ has
been estimated to be on the order of $10 \mu V$ which is three order
of magnitude larger than $10 nV$.

So far we have discussed how to detect the magnitude of the SB.
However, the SB is a vector with its magnitude and spin polarized
direction.\cite{SpinPolar} Its polarized direction can be determined
in the following way. We apply a magnetic field ${\bf B}$ to the
QPC.
%
%
Due to the Zeeman effect the transmission coefficient
$T_{\sigma}(E)$ becomes spin-dependent. Note that $T_{\uparrow}(E)$
is increased and $T_{\downarrow}(E)$ is decreased when ${\bf B}$ is
nonzero, where $\sigma=\uparrow,\downarrow$ represent the parallel
and antiparallel direction of ${\bf B}$, respectively. As a result,
the induced charge current $I_c$ or charge bias $V_c$ is also
affected. By varying the direction of magnetic field ${\bf B}$,
$I_c$ or $V_c$ reaches maximum value when the direction of ${\bf B}$
is parallel to the spin polarized direction of the SB $V_s$.
Therefore the direction of $V_s$ can be determined.

In summary, when an asymmetric spin bias (SB) is applied on a
quantum point contact device or a STM tip, a charge current or
charge bias is induced. The SB can be determined by measuring the
induced charge current or charge bias. Our analysis shows that the
induced charge bias is quite large for almost all parameters and is
well within the reach of the present technology. Hence our proposed
device can efficiently measure the magnitude of the SB by an
electrical measurement. In addition, the spin polarization direction
of the SB can also be measured in the presence of magnetic field.

This work is supported by NSF-China under
Grant Nos. 10525418, 10734110, and 60776060, and a RGC Grant
(No. HKU 7048/06P) from the gov. of HKSAR.


\newpage
\begin{figure}
\caption{(Color online) (a) Schematic plot of an
asymmetric SB applying on a QPC. (b) The transmission coefficient
$T(E)$ vs the energy $E$ for a typical QPC device. In (a) and (b),
the spin-dependent chemical potentials $\mu_{L\sigma}$ are also
shown. (c) Schematics of a suggested device is shown where a FM lead
crossed over an Al lead. A spin current is injected into the Al lead
and a SB is generated along the Al lead. A tip of STM as a detector
to detect the SB.}
\vspace{4cm}
\end{figure}

\begin{figure}
\caption{(Color online) (a) $I_c$ vs $V_g$ at zero
temperature ${\cal T}$ for $V_s=0.005$ (solid line), $0.01$ (dashed
line), and $0.02$ (dotted line). (b) $I_c$ vs $V_g$ at fixed
$V_s=0.02$ for $k_B{\cal T} =0$ (solid line), $0.005$ (dashed line),
and $0.01$ (dotted line). For comparison, $T(E_f)$ vs $V_g$ is also
plotted in (b), [see gray (or red) solid line].}
\end{figure}

\begin{figure}
\caption{(Color online) (a) and (b) are $V_c/V_s$
vs $V_g$ for different SB $V_s$ (a) and different temperature
$k_B{\sl T}$ (b). (c) is $V_c/V_s$ vs $V_s$ for different $V_g$, and
(d) is $V_c/V_s$ vs ${\cal T}$ for different $V_g$. The other
parameters are same to that in Fig.2.}
\vspace{5cm}
\end{figure}

\newpage
$             $
\begin{figure}
\includegraphics
{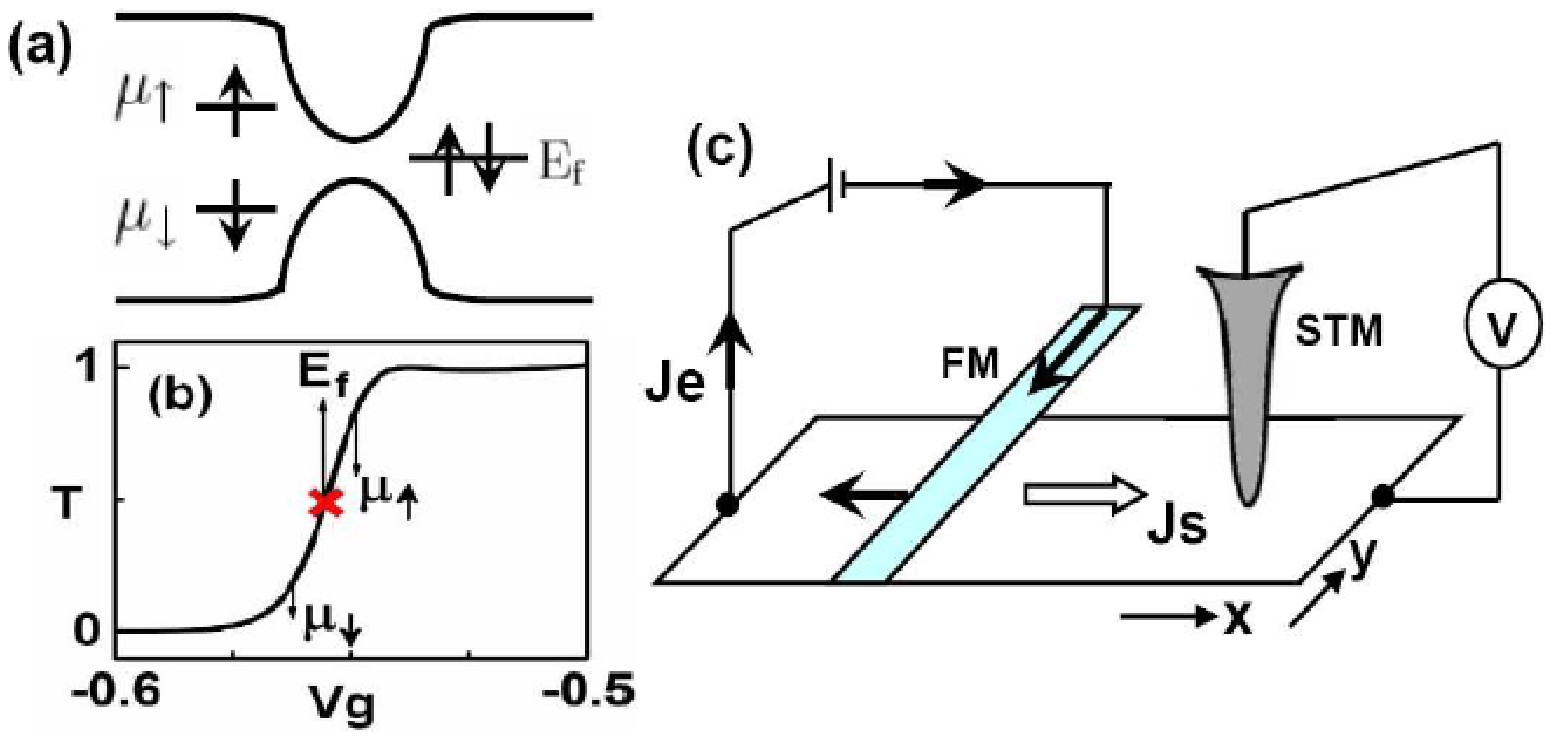}
\end{figure}
$             $

\newpage
$             $
\begin{figure}
\includegraphics
{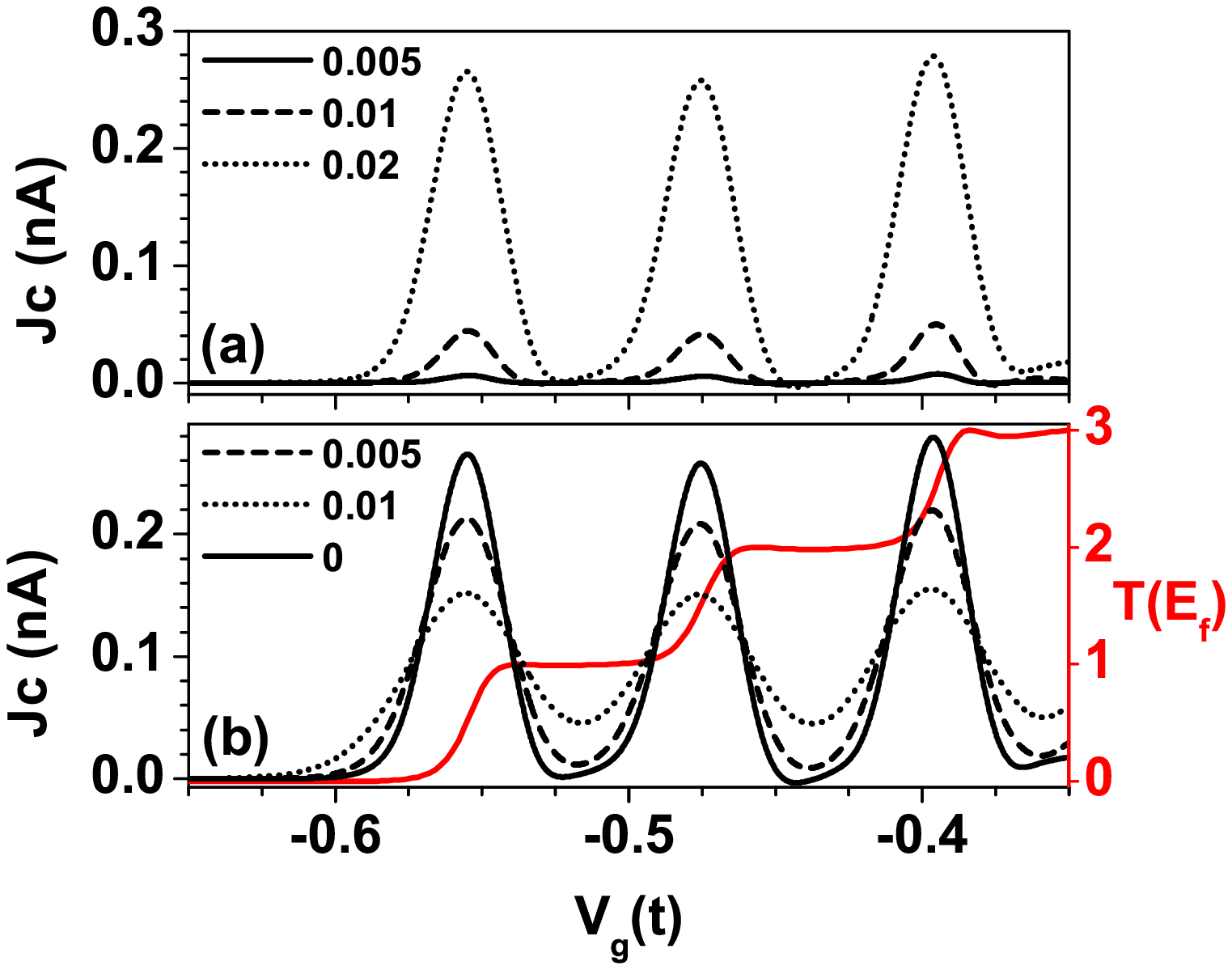}
\end{figure}
$             $
\newpage
$             $
\begin{figure}
\includegraphics
{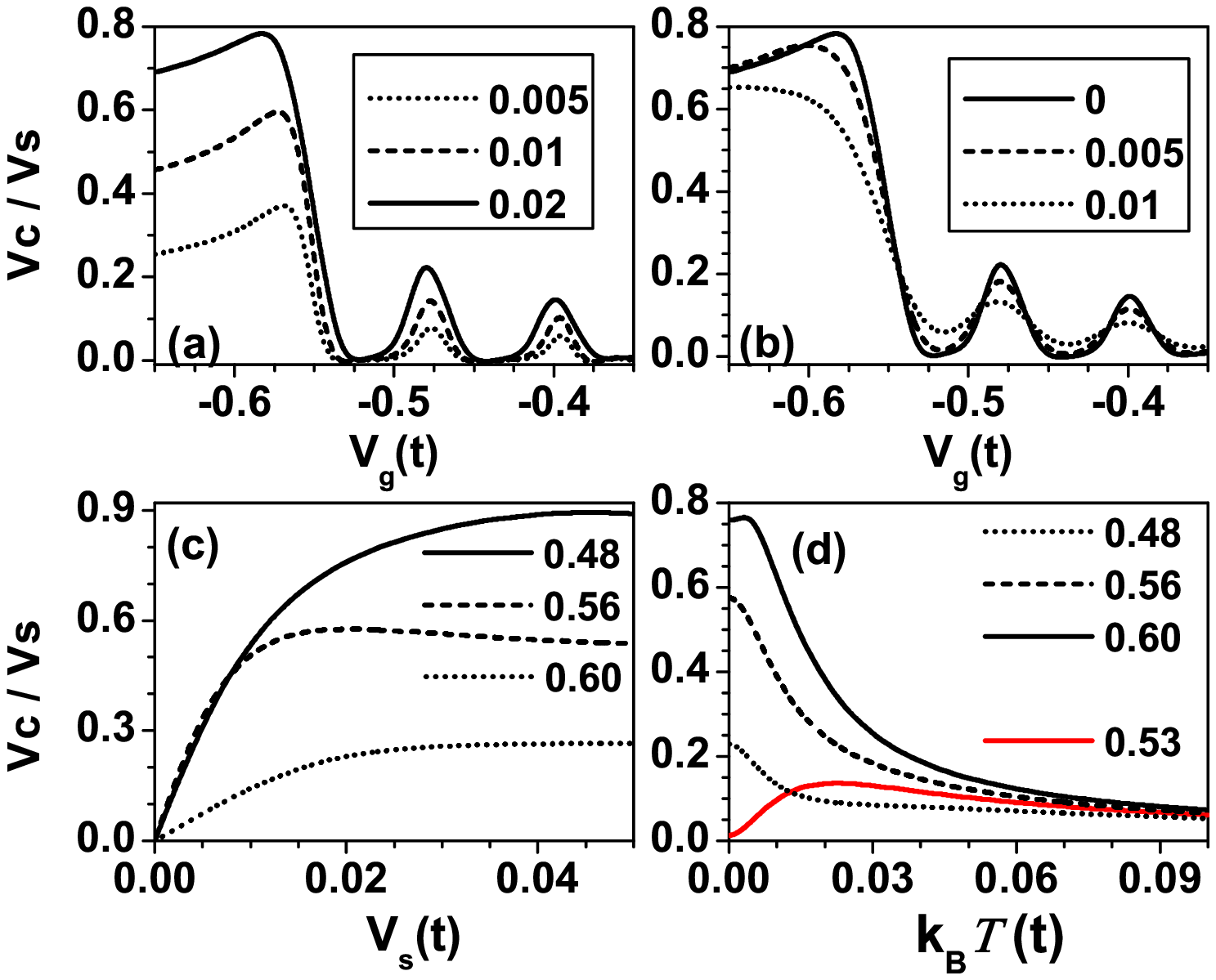}
\end{figure}

\begin{thebibliography}{00}
\bibitem{note1}
The asymmetric SB is crucial in our proposal since we need the
non-zero charge current. If a symmetric SB with $\mu_{L\uparrow}=
-\mu_{L\downarrow}=-\mu_{R\uparrow}= \mu_{R\downarrow}= V_s$ is
applied to the QPC, the induced charge current $I_c$ is always zero
and our method does not apply. In fact, in most cases, the SB is
asymmetric. For example, in the proposed experimental setup in
Fig.1c, the chemical potential in the STM is always
spin-independent, so there is always an asymmetric SB.
\end{thebibliography}
\end{document}